\documentstyle[12 pt]{article}
\begin{document}
\begin{center} {\huge Simulation of Hierarchical Viscous Fingering Pattern
in Lifting Hele-Shaw Cell} \vskip .5 cm Tapati Dutta \\ Physics Department, St Xavier's College\\ Kolkata 700016, India \vskip .3 cm Subrata Kumar Kabiraj and Sujata Tarafdar \footnote{ 
corresponding author, email: sujata@juphys.ernet.in}\\
Condensed Matter Physics Research Centre\\Physics Department, Jadavpur University\\
Kolkata 700032, India \end{center}
\noindent
{\bf Abstract}\\
Viscous fingers in the lifting Hele-Shaw cell form a hierarchical pattern due to competition
between growing fingers in a converging geometry. If the defending fluid is
visco-plastic, a permanent three-dimensional pattern is formed, which is usually approximated 
as quasi-two-dimensional. We present here a Monte-Carlo  simulation which attempts to 
reproduce the real 3-dimensional pattern in rectangular geometry using a very simple algorithm. We study the
finger length distribution, the rate of increase of coverage of the invading fluid, the 
height profile of the final pattern and observe  how these change with characteristics of the
fluids.

\section{Introduction}

Viscous fingering (VF) patterns \cite{rev1,rev2} belong to the class of instabilities which are easily produced experimentally
but quite complex to study analytically. Existing work mostly focusses on the single Saffman-Taylor finger in a rectangular 
Hele-Shaw (HS) cell and details of its stability and shape \cite{mm}. The circular pattern in a lifting Hele-Shaw cell, formed by
competition and interaction between a large number of fingers is however a fascinating problem \cite{ju1,ju2} 
but much less studied.

When a less viscous fluid is forced into a more viscous fluid under pressure, the interface between the two becomes
unstable and the less viscous `invading' fluid (fluid 1) enters the more viscous 
`defending' fluid (fluid 2) in the form of irregular finger
like protrusions which may repeatedly branch and sometimes form a fractal pattern.
The Hele-Shaw cell, consisting of two parallal glass plates, with a very small gap  in between, is the most popular model
for study of  VF patterns. The more viscous  fluid 2, is kept sandwiched between the 
two plates, and the less viscous  fluid 1 is forced in
the gap, through a small hole at the centre of the upper glass plate.
 Fluid 1 forms complex patterns, depending on the characteristic properties of the two fluids.
 The viscosities of the fluids, their interface tension, the rate of flow of the invading fluid are some of the properties
affecting the pattern. Many modifications of this simple HS cell have  been developed.

One such apparatus  is the `lifting Hele-Shaw cell' (LHSC). In this  modification of the conventional Hele-Shaw cell,
the gap between the two plates increases with time. The upper plate is lifted slowly keeping it parallel to the lower
 one. This causes a sucking effect drawing in the lower viscosity fluid, which surrounds the high viscosity fluid sandwiched between the plates. Quite a large number of papers have been published on the LHSC \cite{LHSC1,LHSC2,ju1,ju2}, some theoretical
and numerical studies have also been done \cite{shel}.
If the defending fluid in the LHSC is a visco-plastic paste, the pattern on the two plates after seperation, is a permanent
three-dimensional pattern, it is actually the complement of the fingers, where the displaced fluid has accumulated.
The ridges between two adjacent fingers show a variation in height, and there are peaks at the forks where two ridges
join. The height of the ridges is less at the edges and increases towards the centre of the circular plate. This interesting pattern has been studied experimentally \cite{vanD,ju1} but we do not 
know of any attempt at
 simulating it.

In the present paper we report a Monte-Carlo simulation of VF patterns in a Hele-Shaw cell with a visco-plastic material
such as oil-paint, displaced by a less viscous fluid which may be air or a newtonian liquid. We study characteristics of 
the pattern, such as survival exponent of the fingers growing upto different lengths, rate of increase in coverage of the
displacing fluid with time, and the time required for breakthrough as well as the morphology of the patterns of both the
displacing and the displaced fluid. The displaced fluid being visco-plastic piles up when displaced adding an extra
dimension (both metaphorically and physically) to the problem. All these features change with parameters representing
viscosity contrast, interface tension and density of the fluids. We  compare our results with experiments on a lifting
Hele-Shaw cell (LHSC) using different fluids.

\section{The simulation algorithm}

Growth of viscous fingers and diffusion limited aggregates (DLA) generated by random walks are both examples of Laplacian 
growth \cite{rev1}. So random walk algorithms have been used to simulate viscous fingers. However, this has problems
because such models produce highly branched and irregular fingers similar to surface deposition models \cite{Barab}. Different
methods are used for smoothening the structures to account for the effect of interface 
tension \cite{rev1}. Another problem is
that  there is no 
provision in such models for taking into account the physical properties of the two fluids and observing the changes in
morphology when these properties change.
In our model, we calculate the pressure gradient assuming a linear variation from a constant low pressure isobar, and
assign growth probability in proportion to the pressure gradient, according to Darcy's law \cite{dar}. The proportionality
constant involves a parameter representing the viscosity contrast between the two fluids. We assign another parameter
which depends on the viscoplastic nature of the defending fluid, making the probability of displacement smaller, when a large
amount of it has piled up at a site. In addition there is a parameter representing surface tension which smooths out
indentations in a finger during growth. We find that the different morphological features observed in the LHSC experiment
can be reproduced by controlling these three parameters. We further quantify our results by recording the exponents 
governing power-laws connecting certain features of growth, these are compared with experimental results where possible.

Our pattern is simulated on a rectangular two-dimensional lattice of size $(L_x \times L_y)$. We start with an initial
uniform `mass' distribution of the defending fluid (fluid 2) over the whole system, assigning one
unit of fluid 2 to each site. 
This is represented by $$ m(x,y,t_{=0}) = 1$$ for all (x,y). An initial random disturbance is 
created along one edge at $y=0$ by 
introducing a column of fluid 1 with lengths varying randomly between 1 to 5 units at alternate
sites. Now the fluid 1 columns grow both forward in the y direction and sideways in the x 
direction , with growth probability
determined as follows
\begin{equation} P_{f} = 1/D \mu {\nabla p(x,y,t) m(x,y,t)} \end{equation}
and \begin{equation} P_s = 1/D m(x,y,t) \end{equation}
here $P_f$ and $P_s$ denote respectively the probabilities for forward and lateral growth.
$\nabla p(x,y,t)$ is the pressure gradient at $(x,y)$ at x,y at time t, calculated 
assuming a linear 
variation from a constant lower pressure at the far end $y = L_y$. $\mu$ represents the
viscosity contrast between the two media and $D$ the resistance of fluid 2 to pushing by 
fluid 1 (this maybe related to the yield stress). The growth processes are carried out 
sequentially, the probability determined using two random number generators.

When fluid 1 proceeds by one site in the forward or lateral direction, the `mass' of fluid
2 previously occupying that site is pushed to the next site. So as the fingers of fluid 1
grow, the fluid 2 accumulates forming ridges with peaks at the forks. this mimics the
visco-plastic nature of fluid 2 and the increasing separation between the plates in the
LHSC.
 After completing
one sequence of forward and lateral growth, the indentations formed in the fingers are
smoothed out with a certain probability determined by the surface tension parameter $T$.
Growth processes are repeated until the breakthrough point, when one of the fingers reaches
the other end at $y=L_y$. It may be noted that though we use the term `breakthrough' as in the
normal HS cell, in the LHSC, this is actually the time when the fingers reach the centre and
the plates separate from each other.

 We record the lengths of all the fingers as they grow, the total
coverage by fluid 1 (i.e. the number of sites occupied by fluid 1 as function of time) and 
the  time for breakthrough $t_{br}$. We also record the final distribution of fluid 2, which is
three-dimensional since the visco-plastic displaced fluid is allowed to accumulate 
vertically. So we have a relief map of $m(x,y,t_{br})$. The maximum height $h_{max}$ attained at time $t_{br}$
represents twice the separation of the plates at that time. The factor 2 is due to  the fact that the
same pattern is formed on the upper plate also. In contrast to the conventional Hele
-Shaw cell the total amount of fluid 2 in the cell is conserved.

As we want to study the morphology of a particular pattern, each pattern is generated
separately, without any averaging. But we generate several (about 10) patterns with the same set of
parameters changing the random number sequence in order to get repeated observations of the
exponents and other quantities studied, we report average values of these.

\section{Results of computer simulations}

Some typical fingering patterns generated are shown in figs 1(a)-(d). The different fingers are
shown in different colours for clarity. The hierarchical character of the pattern formed by
termination  of smaller fingers due to the viscous instability is quite clearly seen.
There is a strong similarity to the inverted deterministic binary tree. This is observed in 
many other natural systems, for example river basin boundaries \cite{riv},and crystal growth in rock cavities \cite{crys}.

There is a noticable change in morphology on varying the parameters $\mu, D$ and $T$. 
Decreasing $\mu$ decreases the viscosity contrast and enhances forward growth compared to 
lateral growth. This makes longer and narrower fingers which grow parallelly with much less
competition and hence a slower reduction in number of surviving fingers.

$T$ the surface tension parameter controls the branching of the fingers, when it is zero,
fingers have very little branching and look wide and compact, see fig.1(a). Higher $T$ (here the highest
possible value is $T = 1$), makes the fingers highly ramified as shown in fig. 1(b). 

Table I gives the numerical values characterising  the pattern morphology for different values of the
parameters. We calculate the  survival exponent $f_{surv}$, the breakthrough time $t_{br}$ and the
maximum height reached $h_{max}$.

The height distribution is controlled by $T$ and $D$.
$T$ affects
the height distribution very strongly. For $T=0$ the highly ramified fingers do not allow the fluid 2
to pile up much and the height distribution is more or less uniform. this can be seen from the values of
$h_{max}$ in Table I, for $T=0$ $h_{max}$ is quite small, except when $\mu$ is very low. For very low
$\mu$ the straight and parallel fingers reach the end in the shortest time possible on the lattice, i. e. in $L_y$
units of time, and fluid 2 is also pushed straight along to its maximum possible height, for  $T = 1$.
For $T = 0$ $h_{max}$ is somewhat lower. This can be seen from figures 1(a-d) and 2(a,b).

The effect of $D$ is as follows -
for large  $D$ it is more difficult for fluid 1 to displace fluid 2, after some accumulation,
this causes more branching of the fluid 1 fingers and a more uniform distribution of fluid 2.
If $D$ is small there are higher ridges and peaks of fluid 1 towards the far end of the 
pattern, while fluid 2 fingers look more compact and rounded. These features are exhibited in
experimental patterns as we discuss in a later section.

To quantify the overall characteristics of our patterns we note the following. The number of fingers
surviving upto a certain length as the fingers grow. The cumulative length distribution
shows a linear behaviour in a log-log plot. $log N(l)$ vs. $log(l)$ plot is shown in fig. 3 
where $N(l)$ is the number of fingers with length greater than$l$. The slope of this plot $f_{surv}$ is a
measure of the competition in finger growth. The values of $f_{surv}$  are plotted in table I for different
values of the parameters. 
\begin{equation} N(l) \propto l^{f_{surv}} \end{equation}
The graph shows a stepped behaviour for higher values of$l$, this
is connected with the discrete scale invariance observed in LHSC patterns \cite{ju2}.
The stepped region does not always follow the linear fit for  Log N(l) vs. log(l) very well,
in such cases we report the slope of the initial linear region.

Another feature we study is the increase in total air coverage with time. The coverage $C(t)$, 
defined as the total area occupied by fluid 1 at time $t$ grows linearly with t upto some time
but shows an upward trend later, this is shown in fig.4. However a log-log plot does not show linear behaviour for the
whole time upto breakthrough $t_{br}$. The values of $t_{br}$ for different parameters are
shown in Table I in units of $L_y$. Also shown is $h_{max}$, the maximum height upto which the fluid 2 
accumulates at breakthrough, this is also  in units of $L_y$.

\section{Comparison with Experiments on LHSC}

We have conducted experiments on the LHSC using different fluid combinations. The experimental
setup is described elsewhere in detail \cite{ju1,ju2}, we give a brief description here.
Our lifting Hele-Shaw cell consists of two thick $(\sim 0.5cm)$ glass plates. The lower one is fixed to a rigid frame,
and the upper one can be lifted using a pneumatic cylinder arrangement. The lifting force can be adjusted. In the
 standard LHSC for studying adhesion the velocity of plate separation is constant, whereas in our apparatus the lifting force
is controlled and kept constant during the experiment, allowing the velocity to vary. The process of pattern formation is
recorded by a CCD camera and analysed using image-pro plus software.

The defending fluid 2 is placed on the lower plate, and the upper plate pressed down on it. This makes the fluid 2  form a circular blob of 2-3 cm. diameter. Then the upper plate is slowly lifted, allowing air to enter from the sides, while recording the pattern formation from below the lower
plate. If the invading fluid is not air, the appropriate fluid is placed surrounding the defending fluid, taking care that
no air bubbles are left in the gap. Formation of the final pattern with the two plates no longer touching takes typically 20-50 seconds.
An identical replica of the pattern is formed on the upper plate as well, due to the symmetry of the arrangement, as the
effect of gravity is negligible here. The height of the highest peak on the pattern is therefore, half of the separation
between the plates at breakthrough.

In the experiments we find the behaviour of the fingering pattern is qualitatively
similar and the range of the characteristic exponents from experiment and computer simulation
match quite well. The survival of fingers shows a power law behaviour on the average with
a linear log-log plot on which
oscillations or steps are superposed, (see fig. 5a-b). For the high viscosity contrast case (fig.5a) where fluid 1 is air
and fluid 2 oil-paint, the slope $f_{surv}(exp)$ of the log-log plot of  $N(l)$ vs.$ l$ is very close to  1,
whereas for water-paint combination (fig. 5b) where viscosity contrast is lower the slope is less
than 1. Appearance of the patterns is also well reproduced. Experiments with other fluids are in progress. Further details 
of these  experiments
will be reprted elsewhere.
 \cite{ju3}.

As we have used a rectangular lattice for simplicity, we cannot expect the pattern generated
to match experiments completely. In the last stages of pattern formation close to
breakthrough the converging circular geometry in the experiment will lead to results very
different from what would be produced in a  linear cell. But for a blob of fluid 2 with
an initial radius  large compared to the finger width initially the pattern of growth will
be simiar to the linear case. Of course, linear geometry is not possible in a LHSC exeriment
with parallel plate separation.

\section{Conclusion}

The present work is an attempt to generate by computer simulation the hierarchical pattern
of viscous fingers observed in the lifting Hele-Shaw cell. Different characteristics and their variation with properties of the two fluids is presented, the three dimensional nature of
the pattern in a viscoplastic defending fluid is taken into account. Predominance of power law
behaviour in finger length distribution is oberved and the survival exponent $f_{surv}$ takes values ranging from
$\sim 0.5$ to $\sim 1.5$ for increasing viscosity contrast. Variation in height of the
pattern is controlled by another parameter $D$ characterising the visco-plastic behaviour of the defending fluid. $D$ also controls the ramification to some extent, large $D$ leads to
some branching. However, branching is principally determined by the surface tension parameter
$T$.
Oscillations about the power law are indicative of discrete scale invariance
\cite{dsi,ju3} we propose to investigate this aspect in detail later. Our simulation results
are compared with experiments on LHSC patterns generated in our laboratory and we find 
qualitative as well as quantitative agreement.

We have not yet systematically done an experimental investigation of LHSC patterns, varying
the interface tension or yield stress of the fluids, but further work is in progress.

In the present work we have assumed Darcy's law to be valid, not taking into account the
non-Newtonian nature of visco-plastic paste like materials, neither do we consider explicitly
the effect of changing plate separation. this should introduce a time-dependent parameter
in Darcy's law. We hope to develop  a more realistic algorithm in future as the preliminary
results are encouraging. 
\section{Acknowledgement}
The authors thank Martine Ben Amar and Yves Couder for helpful discussion. Indo-French Centre
for the Promotion of Advanced Research/ Centre Franco-Indien Pour le Promotion de la 
Recherche Advancee is  gratefully acknowledged for grant
of a research project. S N Bose National Centre for Basic Sciences is acknowledged for
extending computer facilities.

\newpage \noindent Table I. Variation of characteristics of the VF patterns with fluid properties.
\vskip .5 cm

\begin{tabular}{|c|c|c|c|c|c|}
\hline $\mu$ & T &D& $f_{surv}$ & $T_{br}/L_y$ &$h_{max}/L_y$\\ \hline
0.1 & 1.0 &1.0 & -1.466 &1.785 &0.51 \\
0.1 & 1.0 &20 & -1.380 & 34.56 &0.96\\
0.1& 0.0 & 1.0 &-1.086 & 1.93 & 0.16\\
\hline
0.02 & 1.0 & 1.0 & -1.20 & 1.04 & 0.77\\
0.02 & 1.0 & 5.0 & -1.09 & 2.78 & 0.56 \\
0.02 & 0.0 & 5.0 & -1.05 & 3.42 & 0.14\\
\hline 
0.002 & 1.0 & 1.0 & -0.652 & 1.0 & 1.0\\
0.002 & 1.0 & 5.0 & -0.733 & 1.077&0.66\\
0.002 & 0.0 & 1.0 & -0.576 & 1.0&0.49\\
\hline
0.0002 & 1.0 & 5.0& -0.484 & 1.0 & 1.0\\
0.0002&0.5 & 5.0 & -0.502 & 1.0 & 1.0\\
0.0002 & 0.0 & 5.0 & -0.499 & 1.0& 0.66\\

\hline      
\end{tabular}
\newpage
\noindent {\bf Figure Captions}

Figures 1-2 available on request. /vskip .25 cm
\vskip .25 cm
Figure 1a. Finger patterns for high viscosity contrast, $\mu = 0.1$ and large interface tension $T = 1$. The fingers 
 (of fluid 1) are shown in colour and accumulated fluid 2 in white.

\vskip .25 cm
Figure 1b. Finger patterns for high viscosity contrast, $\mu = 0.1$ and small interface tension $T = 0$. The fingers
 (of fluid 1) are shown in colour and accumulated fluid 2 in white.

\vskip .25 cm
Figure 1c. Finger patterns for low viscosity contrast, $\mu = 0.001$ and large  interface tension $T = 1$.
The value of $D$ is low here.  The fingers
 (of fluid 1) are shown in colour and accumulated fluid 2 in white.

\vskip .25 cm
Figure 1d. Finger patterns for low viscosity contrast, $\mu = 0.001$ and large  interface tension $T = 1$. 
The value of $D$ is high here.  The fingers
 (of fluid 1) are shown in colour and accumulated fluid 2 in white.

\vskip .25 cm
Figure 2a. Three-dimensional representation of the fingering pattern showing accumlation of fluid 2 in
ridges, for $T = 0$. The height distribution of fluid 2 is more or less uniform.

\vskip .25 cm
Figure 2b. Three-dimensional representation of the fingering pattern showing accumlation of fluid 2 in
ridges, for $T = 1$. The height distribution of fluid 2 shows strong variation and fingers of fluid 1 are
wide and compact.

\vskip .25 cm
Figure 3. Log-log plot of $N(l)$ - the number of surviving fingers upto length $l$ versus l

\vskip .25 cm

Figure 4. The variation in coverage of fluid 1 versus time.

\vskip .25 cm
Figure 5a. Experimental results for number of surviving fingers $ log N(l)$ vs $log l$, for high viscosity contrast
with air as fluid 1 and oil-paint as fluid 2. Slope of the linear fit $f_{surv}(exp)$ is -0.99.

\vskip .25 cm
Figure 5b. Experimental results for number of surviving fingers $ log N(l)$ vs $log l$, for low viscosity contrast
with water as fluid 1 and oil-paint as fluid 2. Slope of the linear fit $f_{surv}(exp)$ is -0.77.

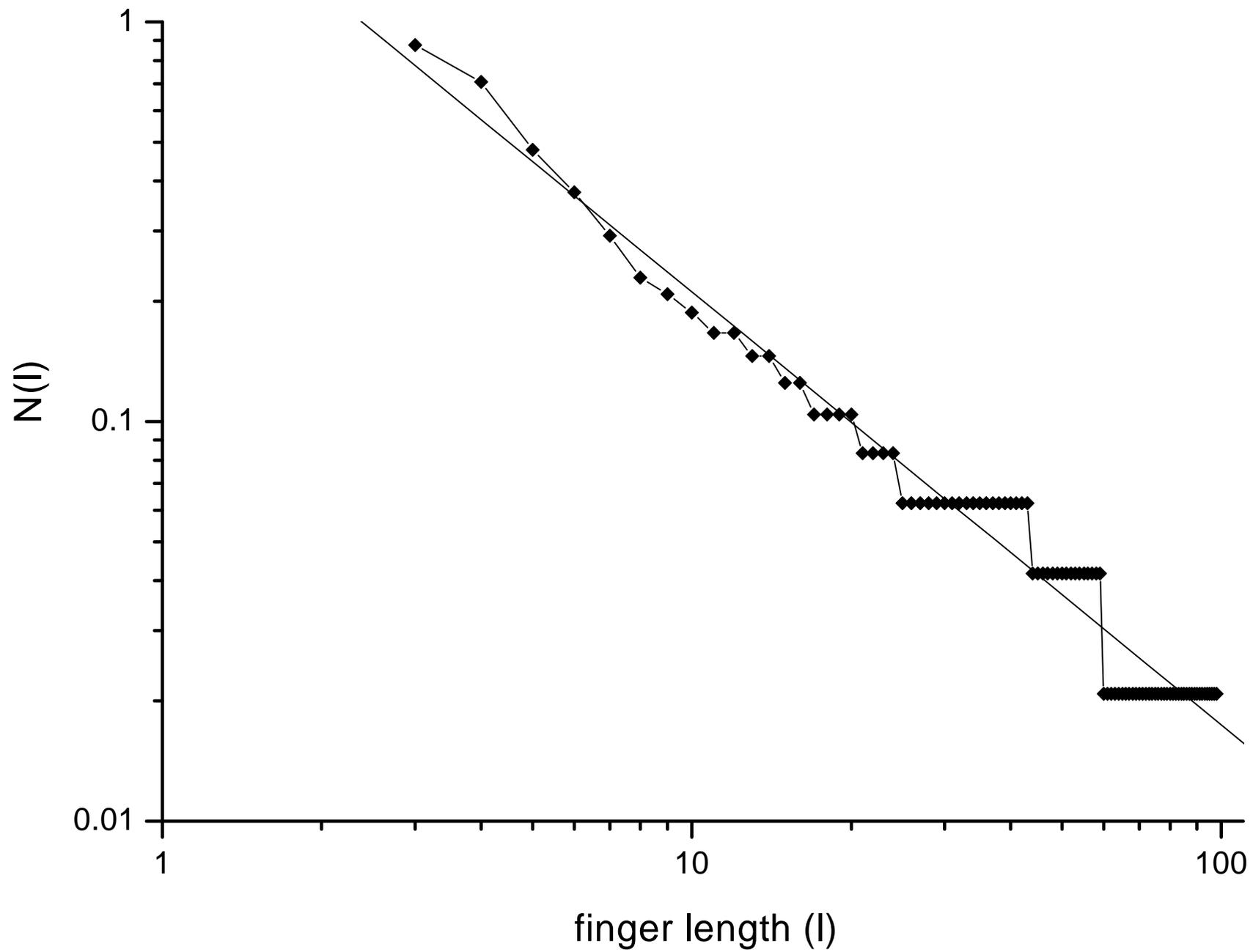
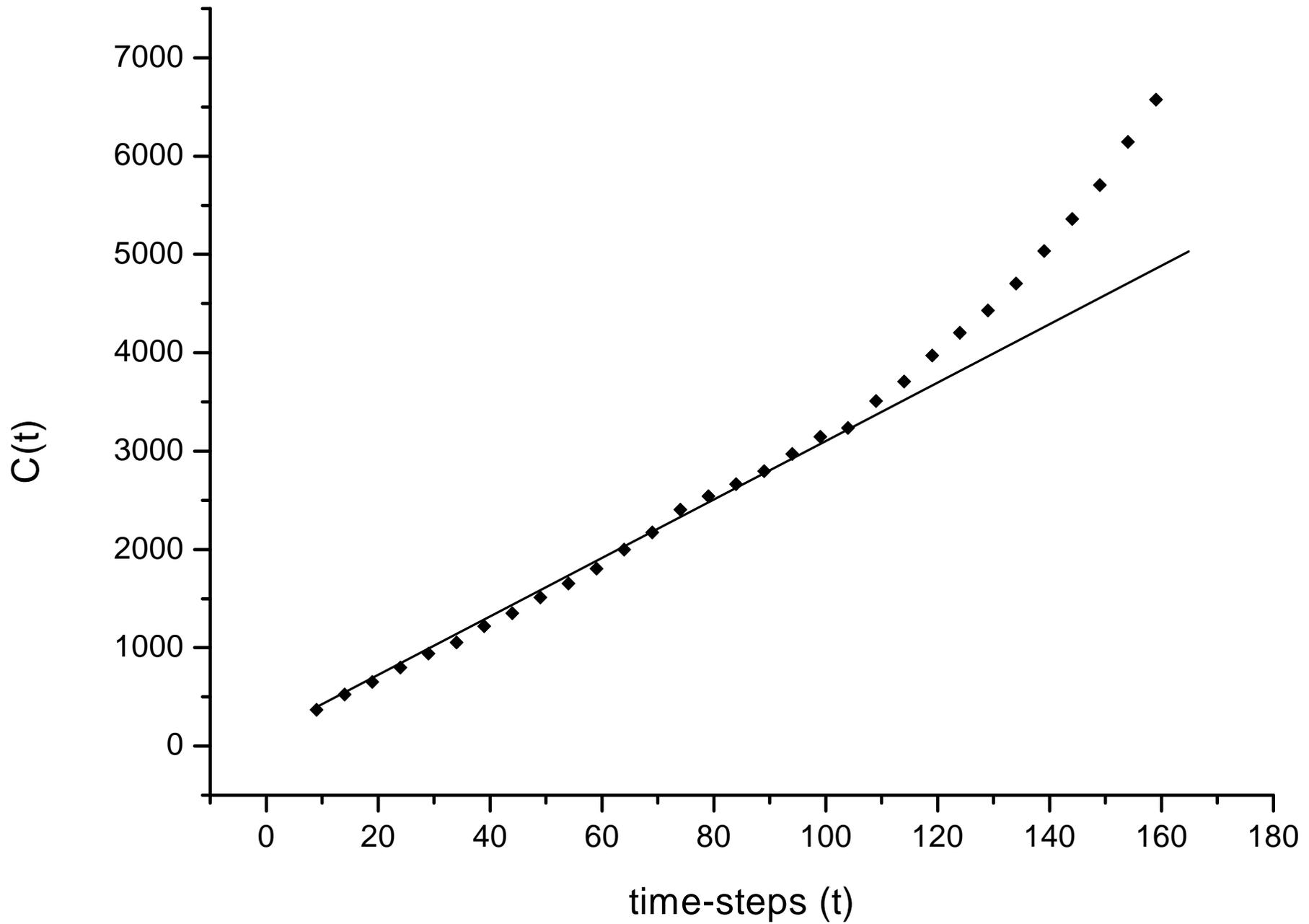
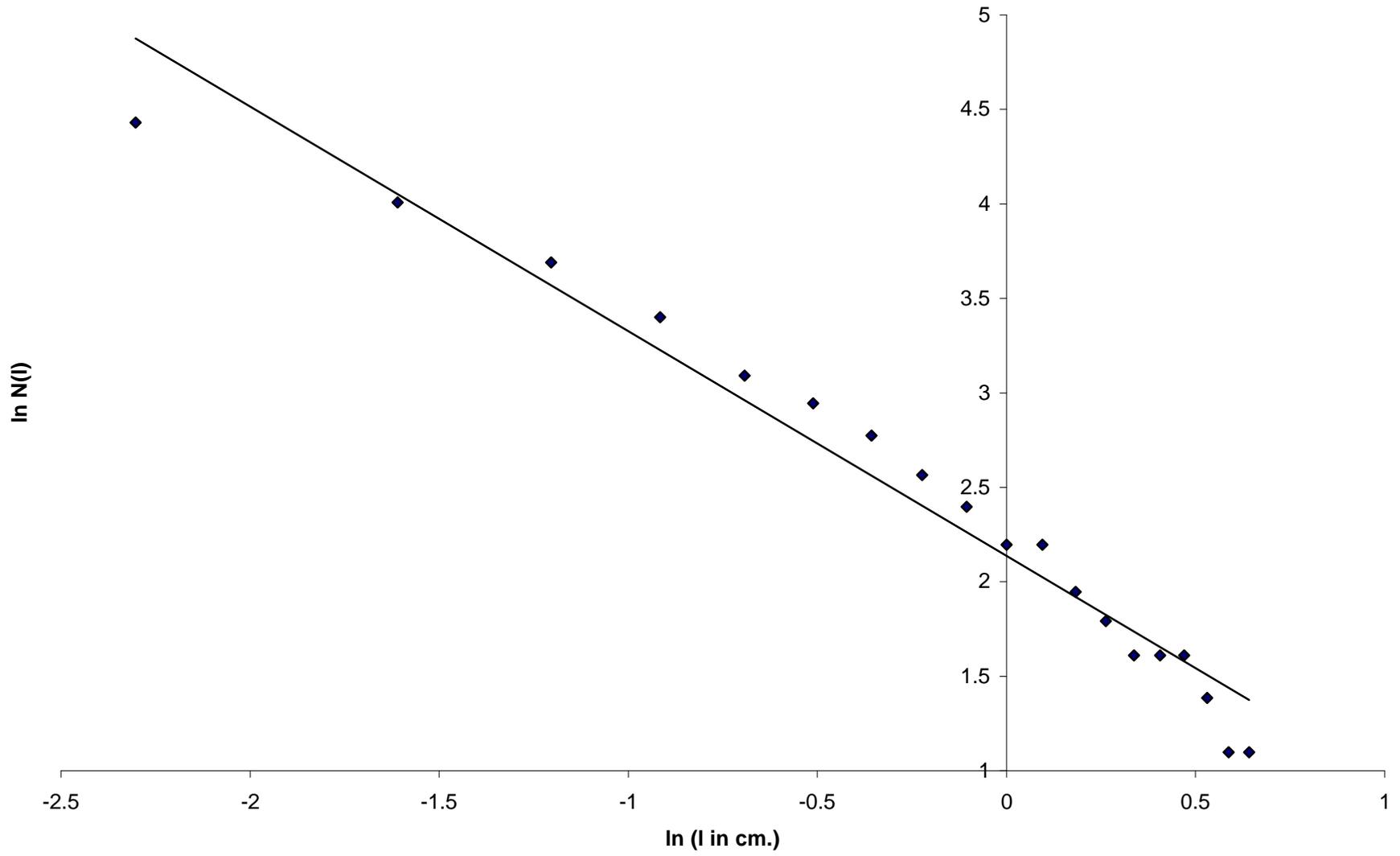
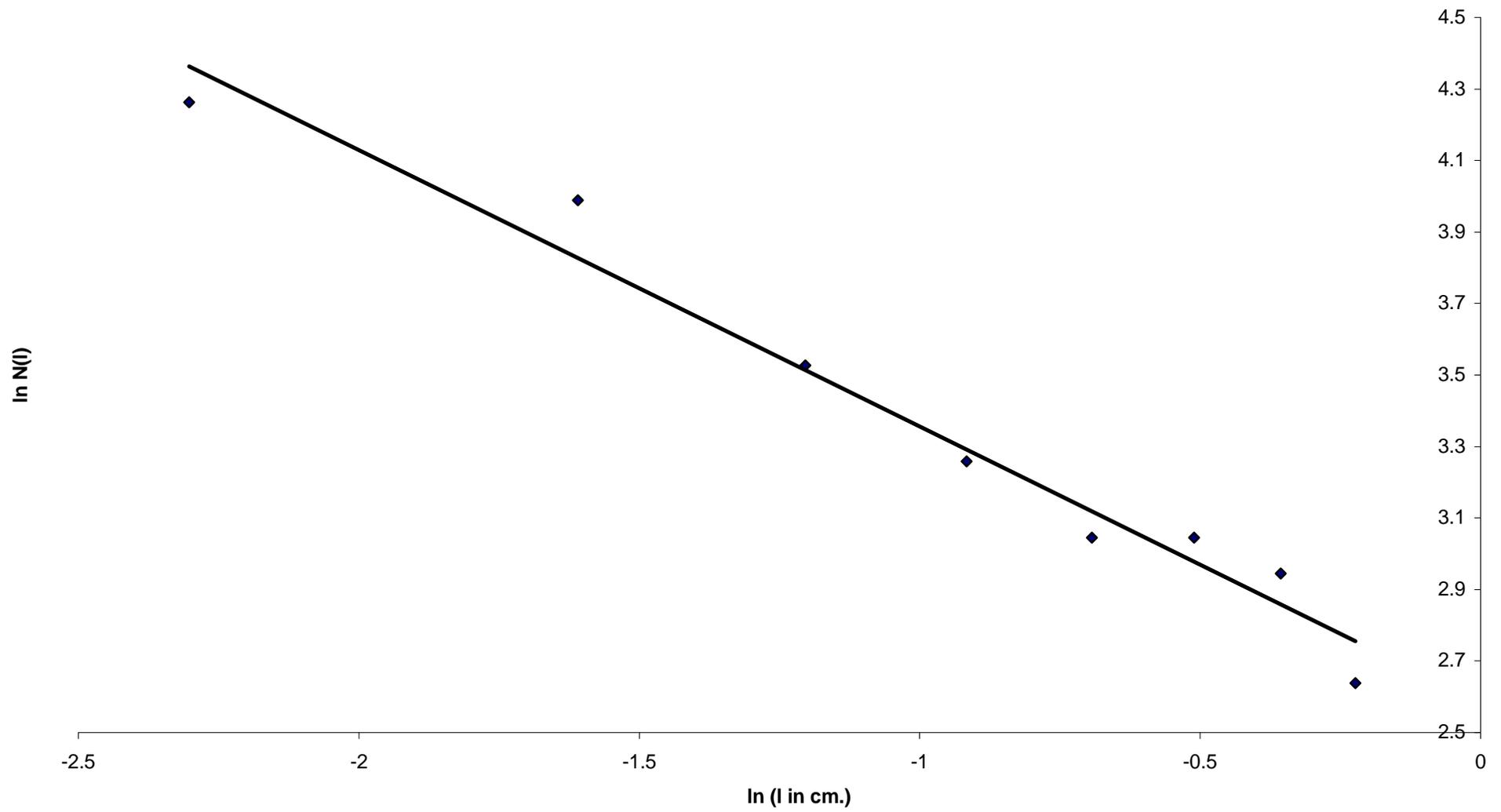
\end{document}